\documentclass[aps,prl,twocolumn,showpacs,a4paper,superscriptaddress]{revtex4}
\usepackage{graphicx}
\usepackage{amsmath}
\begin{document}
\title{Electronic structure of epitaxial graphene layers on SiC: effect of the substrate.}
\author{F.Varchon}
\affiliation{Institut N\'eel, CNRS-UJF, BP 166, 38042 Grenoble Cedex 9 France\\}

\author{R.Feng}
\affiliation{The Georgia Institute of Technology, Atlanta, Georgia
30332-0430, USA\\}

\author{J.Hass}
 \affiliation{The Georgia Institute of Technology, Atlanta, Georgia
30332-0430, USA\\}

\author{X.Li}
 \affiliation{The Georgia Institute of Technology, Atlanta, Georgia
30332-0430, USA\\}

\author{B.Ngoc Nguyen}
\affiliation{Institut N\'eel, CNRS-UJF, BP 166, 38042 Grenoble Cedex 9 France\\}

\author{ C.Naud}
 \affiliation{Institut N\'eel, CNRS-UJF, BP 166, 38042 Grenoble Cedex 9 France\\}

\author{P.Mallet}
\affiliation{Institut N\'eel, CNRS-UJF, BP 166, 38042 Grenoble Cedex 9 France\\}

\author{J.-Y.Veuillen}
\affiliation{Institut N\'eel, CNRS-UJF, BP 166, 38042 Grenoble Cedex 9 France\\}

\author{C.Berger}
\affiliation{Institut N\'eel, CNRS-UJF, BP 166, 38042 Grenoble Cedex 9 France\\}

\author{E.H.Conrad}
\affiliation{The Georgia Institute of Technology, Atlanta, Georgia
30332-0430, USA\\}

\author{L.Magaud}
\affiliation{Institut N\'eel, CNRS-UJF, BP 166, 38042 Grenoble Cedex 9 France\\}

\date{\today}

\begin{abstract}
Recent transport measurements on thin graphite films grown on SiC show
large coherence lengths and anomalous integer quantum Hall effects
expected for isolated graphene sheets. This is the case eventhough
 the layer-substrate epitaxy of these films implies a
strong interface bond that should induce perturbations in the
graphene electronic structure. Our DFT calculations confirm this
strong substrate-graphite bond in the first adsorbed carbon
layer that prevents any graphitic electronic properties for this
 layer. However, the graphitic nature of the film is
recovered by the second and third absorbed layers. This effect is
seen in both the (0001)and $(000\bar{1})$ 4H SiC surfaces. We also
present evidence of a charge transfer that depends on the
interface geometry.  It causes the graphene to be doped and gives
rise to a gap opening at the Dirac point after 3 carbon layers
are deposited in agreement with recent ARPES experiments (T.Ohta
et al, Science {\bf 313} (2006) 951).
\end{abstract}


\maketitle

The possibility of carbon nanotubes (CNT) switching devices has
been pursued in the last decade because of their attractive
electronic properties.  Nevertheless, problems with large
intrinsic resistance in contacts and the inability to control tube
helicity, and thus whether or not they are metallic or
semiconducting, have made large scale circuit designs problematic.
The proposed solution to these problems is an all carbon
nanoelectronics paradigm based on the planar 2D form of carbon,
graphene.\cite{berger2}

Graphene consists of a single carbon plane arranged on a honeycomb
lattice. From a fundamental point of view, graphene ribbons can be
seen as an unrolled CNT but with different boundary conditions
(finite versus cyclic). Therefore, their electronic properties
should be similar. In fact this has been demonstrated in recent
experiments on single and multi-graphene sheets that show the
existence of Dirac Fermions, large electron coherence lengths and
anomalous integer quantum Hall effect ~\cite{berger,
novoselov,zhang}. The advantage of graphene over CNTs for
electronics resides in its planar 2D structure that enables circuit
design with standard lithography techniques. This enables the
graphene to be cut with different shapes and selected edge
direction. By simply selecting the ribbon edge direction it is
possible to design metallic or semiconductor graphene ribbons
\cite{edge,edge2} (analogous to helicity in CNTs).

Since single or multiple sheets must be supported on a surface for
fabrication, the pressing question becomes: how does the interface
between a graphene sheet and its support affect its electronic
properties? In other words can the symmetry of an isolated graphene
sheet be maintained in the presence of an interface? It is this
question that is the focus of this paper. Specifically we have
studied the system of graphite grown on both polar faces of
hexagonal SiC. 

The graphene layers are produced by sublimating Si
from either the 4H- or 6H-SiC (0001) (Si terminated) or
$(000\bar{1})$ (C terminated) surfaces at sufficiently high
temperatures to graphitize the excess carbon \!\cite{forbeaux,
berger2}. 2D transport measurements on these graphitized surfaces
show the presence of Dirac electrons similar to those found on
exfoliated graphene ~\cite{berger,novoselov,rollings, sadowski}.
Besides being a more practical and scalable approach to 2D
graphene electronics, this system has the experimental advantage
of having a well defined interface that can be characterized in
contrast to mechanically exfoliated graphene flakes
~\cite{novoselov, zhang} that must in any case still be supported on a surface
(usually $\text{SiO}_2$).

In this letter, we conclusively show that the first carbon layer
grown above the SiC substrate has no graphitic electronic
properties and acts as a buffer layer between the substrate and
subsequent graphene layers. Atoms in this plane form strong
covalent bonds with the SiC substrate.  The next graphene layer
above the buffer layer shows a graphene-like Dirac band structure
expected for an isolated graphene sheet. The
calculated results are consistent with a short
C first plane-4HSiC$(000\bar{1})$ substrate bond as determined by
Surface X-ray Reflectivity.  Under some conditions, charge
transfer from the substrate results in a n type doping of the graphene layers (Fermi level above the Dirac point). This opens a gap in the graphene bilayer
Dirac bands, in agreement with recent ARPES results \!\cite{ohta}.
Dangling bond related states are found in all tested geometries.
These states can interact with the graphene derived state
depending on the geometry. Their effects on the electronic and
transport properties have to be considered as well as those of the
intrinsic defects of the isolated graphene layer \!\cite{guinea}.

\begin{figure}
\includegraphics[angle=-90, width=6.0cm,clip]{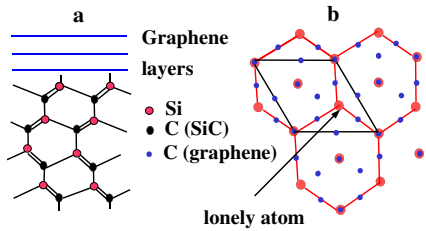}
\caption{(color on line) interface geometry. a- side view, b- top view of the
$\sqrt{3}\times \sqrt{3}R30$ cell in the case of a Si-terminated
SiC face. The lonely atom is missing in the C-deficient geometry}
\label{f.1}
\end{figure}

The systems theoretically studied here are made of one, two or
three carbon layers (on a  honeycomb lattice with Bernal stacking) 
on top of either a SiC 4H (0001) or
$(000\bar{1})$ (Si and C terminated) substrates. Graphene is
nearly commensurate with these SiC surfaces with a common cell
corresponding to a $6\sqrt{3}\!\times\! 6\sqrt{3}R30$ (with
respect to SiC $1\times 1$ surface cell) ~\cite{forbeaux}. This
cell is too large to make realistic calculations. Even the next
smallest nearest commensurate structure, $4\!\times\! 4$ cell (not
experimentally observed), is too large for reasonable
calculations. Therefore, as a first approximation to the actual
structure, we used the $\sqrt{3}\!\times\! \sqrt{3}R30$
reconstruction shown in Fig.~\ref{f.1}. The $\sqrt{3}\!\times\!
\sqrt{3}R30$ cell corresponds to a $2\!\times\! 2$ graphene cell. The
graphene and SiC lattice parameter mismatch requires an $8\%$
stretch of the graphene to make the two cells commensurate. We have
checked that this expansion has no qualitative effect on a free
standing graphene electronic structure (it will however change the Dirac
electron velocity). The cell contains 3 atoms/layer in SiC. At the
interface, two of these atoms are immediately below a C atom in
the first C layer. The third atom (subsequently referred to
as the "lonely atom") has no C atom above it.  A bulk truncated
4H-SiC geometry was used on both faces. We have also checked
another possible interface geometry based on surface X-ray
scattering data:  a C-terminated surface with one C atom out of
three missing (referred to as "C-deficient"). In the bulk
truncated geometry, the lonely atom exhibits a dangling bond (DB)
that points towards the graphene layers. The lonely atom is
suppressed in the C-deficient geometry, thus creating 3 dangling
bonds at the interface.

The electronic structure was investigated using the VASP code
~\cite{vasp}. It is based on Density functional theory (DFT)
within the generalized gradient approximation \!\cite{pw}.  The 4H
SiC substrate is described with a slab that contains 8 SiC
bi-layers with H saturated dangling bonds on the second surface.
The empty space ranges from 15 to 25 \!\AA. Ultra soft
pseudopotentials \!\cite{uspp} are used with a plane wave basis
cutoff equal to 211 \!eV. The experimental graphene layer spacing
was first chosen as the starting value and then all the atoms were
 allowed to relax. Since DFT is known to poorly
describe Van der Waals forces, the final graphene layer
spacing are significantly larger than in the bulk. However, we
point out that the C-short ultrasoft pseudopotential used here has
been extensively tested ~\cite{incze} and was shown to correctly
describe the band structure of graphite in spite of the larger
layer spacings \cite{vdW,evc}. Integration over the Brillouin zone is
performed on a 9x9x1 grid in the Monckhorst-Pack scheme to ensure
convergence of the Kohn Sham eigenvalues. The K point is included
since it is crucial to a good description of the Fermi level for a
single graphene layer.

The X-ray experiments were performed at the Advanced Photon
Source, Argonne National Laboratory, on the 6IDC-$\mu$CAT Ultra
High Vacuum scattering chamber.  The C-Face samples were
graphitized in a vacuum RF-induction furnace ($P\!=
\!3\times10^{-5}{\text{Torr}}$) \!\cite{x_ray} and transported to
the scattering chamber for analysis.

The calculations show that the relaxed geometry of the bulk layers
is influenced only by the first carbon layer. Neither the bulk nor
the first C layer are altered when subsequent C
layers are added. In the last bilayer of the Si (C) terminated
face the lonely atom is displaced toward the bulk by 0.3 \!\AA\!
(0.45), while the bilayer width remains globally unchanged (0.65
(0.7) compared to 0.62-0.63 in the bulk). The first graphene layer
lies 2.0 (1.66) \!\AA\!  above the two outermost atoms. The second
graphitic C plane is 3.8 \!\AA\! (3.9) above the first one
(subsequent planes are spaced by 3.9 \!\AA). As mentionned above, the large value of
the graphene interlayer spacings is due to the known difficulty
of the DFT to describe the van der Waals force \cite{vdW,evc}.  From these results,
we deduce that the interface
carbon layer strongly interacts with the SiC substrate for both Si
and C terminated surfaces. Subsequent C-planes on the other hand
are weakly bound by van der Waals forces as expected for graphite.
This conclusion also holds for the C-deficient geometry. The bulk
relaxation in this latter case is very similar to those of the C
terminated bulk truncated geometry.

\begin{figure}
\includegraphics[width=7.5cm,clip]{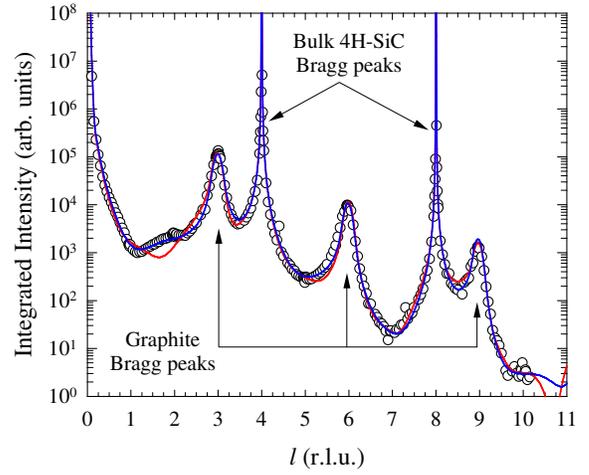}
\caption{(color on line)X-ray $(00l)$ reflectivity data from ~9 graphene layers
grown on the 4H-SiC$(000\bar{1})$ surface. Bulk and graphite Bragg
peaks are labelled. Blue line is the best fit structure with one
reconstructed  SiC bi-layer as described in the text. Red line is
a fit with an extended reconstruction of two SiC bi-layers.}
\label{f.2}
\end{figure}

X-ray reflectivity data confirms this result.  Figure.~\ref{f.2}
shows an experimental reflectivity from $\sim\!9$ graphene layer
film grown on the 4H-SiC$(000\bar{1})$ surface.  Data is on the
$(00l)$ rod in units of $2\pi/a$, where $a\!=\!10.081\text{\AA}$.
Peaks at $l\!=\!4$ and 8 are SiC Bragg reflections, while peaks at
$l\sim$3, 6 and 9 are graphite Bragg points.  A full fit to the
data including substrate relaxation and a multilayered graphite
film is shown.  Details of the fit are given in a separate
article \!\cite{X-ray2}. The model consists of a single
reconstructed SiC bi-layer interface between the graphite and the
bulk. The fit reveals that the first graphene layer is $1.65\pm
.05\text{\AA}$ above the last bulk C-layer consistent with the
calculated value. The next graphene layer is separated from the
first by $3.51\pm 0.1\text{\AA}$ (slightly larger than the bulk
value $3.354\text{\AA}$ \cite{Bulk_C}). Subsequent layers have a
mean spacing of $3.370\pm 0.005\text{\AA}$.  This slightly larger
layer spacing is consistent with stacking faults in the layers
\cite{Faults}. The x-ray results confirm the calculated structure
of a strongly bonded first graphitic layer with a well isolated
graphene layer above it. We note that the extended diamond interface phase
conjectured by others \cite{forbeaux,vanBommel75} does not fit the
x-ray data for the C-face.  This is demonstrated in Fig.~\ref{f.2}
were we force a second SiC bi-layer to be Si depleted by $25\%$.
This fit is obviously worse than a single bi-layer and proves that
the SiC interface is narrow and not extended.

\begin{figure}
\includegraphics[angle=-90, width=11.0cm,clip]{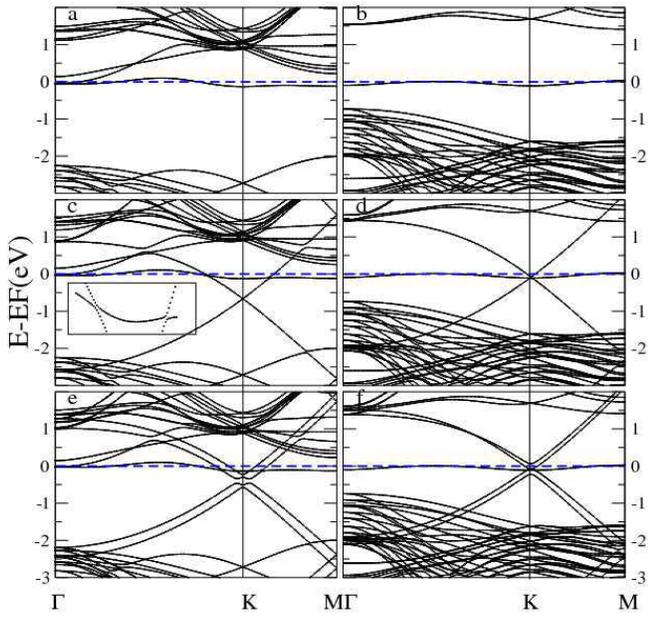}
\caption{(color on line) dispersion curves for one (a,b), two (c,d) and three
(e,f) C layers on bulk truncated SiC. a,c,e correspond to
Si terminated face; b,d,f to C terminated face. Inset in c shows a
zoom of the anticrossing in the vicinity of $E_F$.} \label{f.3}
\end{figure}

The band structures with one (Fig.~\ref{f.3}(a) and (b)), two
(Fig.~\ref{f.3}(c) and (d)) and three (Fig.~\ref{f.3}(e) and (f))
carbon layers on bulk terminated SiC are shown in Fig.~\ref{f.3}. For both polarities, the electronic
structure with a single C layer significantly differs from
graphite ~\cite{mattausch}. It exhibits a large gap and a
Fermi level pinned by a
state with a small dispersion (close to the conduction band for
the Si terminated surface or in the gap for the C terminated
surface). These states are related to the Dangling Bond, DB, of
the lonely atom in the SiC interface layer (a Si (C) dangling bond
state for the Si (C) terminated surface). They remain unchanged when further C layers are added on top of the first one. Figures.~\ref{f.3}(c-f) show that graphene related dispersions are recovered when more than one C layer
is present. In fact, the first C plane acts as a buffer layer that allows growth of subsequent graphene like layers. Indeed one can clearly see for two C planes (buffer+1) the linear dispersion and Dirac
point that are characteristic of an isolated graphene layer (Fig.~\ref{f.3}(c-d)). When three C layers are present (buffer+2) the dispersion is similar to the dispersion of a graphene bilayer ~\cite{latil}.

\begin{figure}
\includegraphics[angle=-90,width=5.5cm,clip]{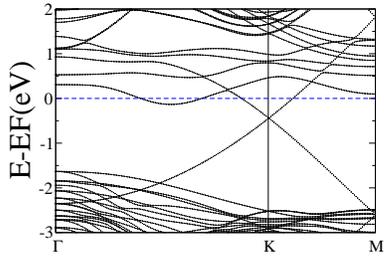}
\caption{(color on line) dispersion curves for 2 C layers on top of the
C-deficient surface} \label{f.4}
\end{figure}

On Si-terminated surface, the Fermi level falls 0.4 eV above the Dirac point. The graphene like planes are n doped. This is confirmed by the opening of a gap in the case of 3 C layers (buffer +2). Tight binding calculations involving $p_z$ orbitals show that this is characteristic of a graphene bilayer where the two planes are not symmetric. In our {\it ab initio} calculation, the Fermi level falls above the highest unoccupied $\pi^*$ band minimum at K point. The comparison to tight binding calculations, shows that the two graphene layers are doped and that one plane is less doped than the other one. This
is in agreement with recent ARPES and XPS measurements ~\cite{ohta,rollings}. For C terminated surfaces in the bulk truncated geometry, the Fermi level falls on the Dirac point and the graphene layers are neutral. On the other hand, the Fermi
level of the C-deficient geometry (Fig.~\ref{f.4}) is 0.4 eV above the Dirac point. It is
fixed by states related to the 3 DB present in this structure. This stresses the role played by interface defects.

For the Si terminated surface the DB related state and the graphene derived
states anticross indicating some interactions between them
(inset of Fig. 3c). This
is not the case for the bulk C terminated surface. This effect may
have a crucial impact on transport properties of the film and
explain the low electron mobilities of Si-face films compared to
C-face films.\cite{x_ray} 

\begin{figure}
\includegraphics[width=7.0cm,clip]{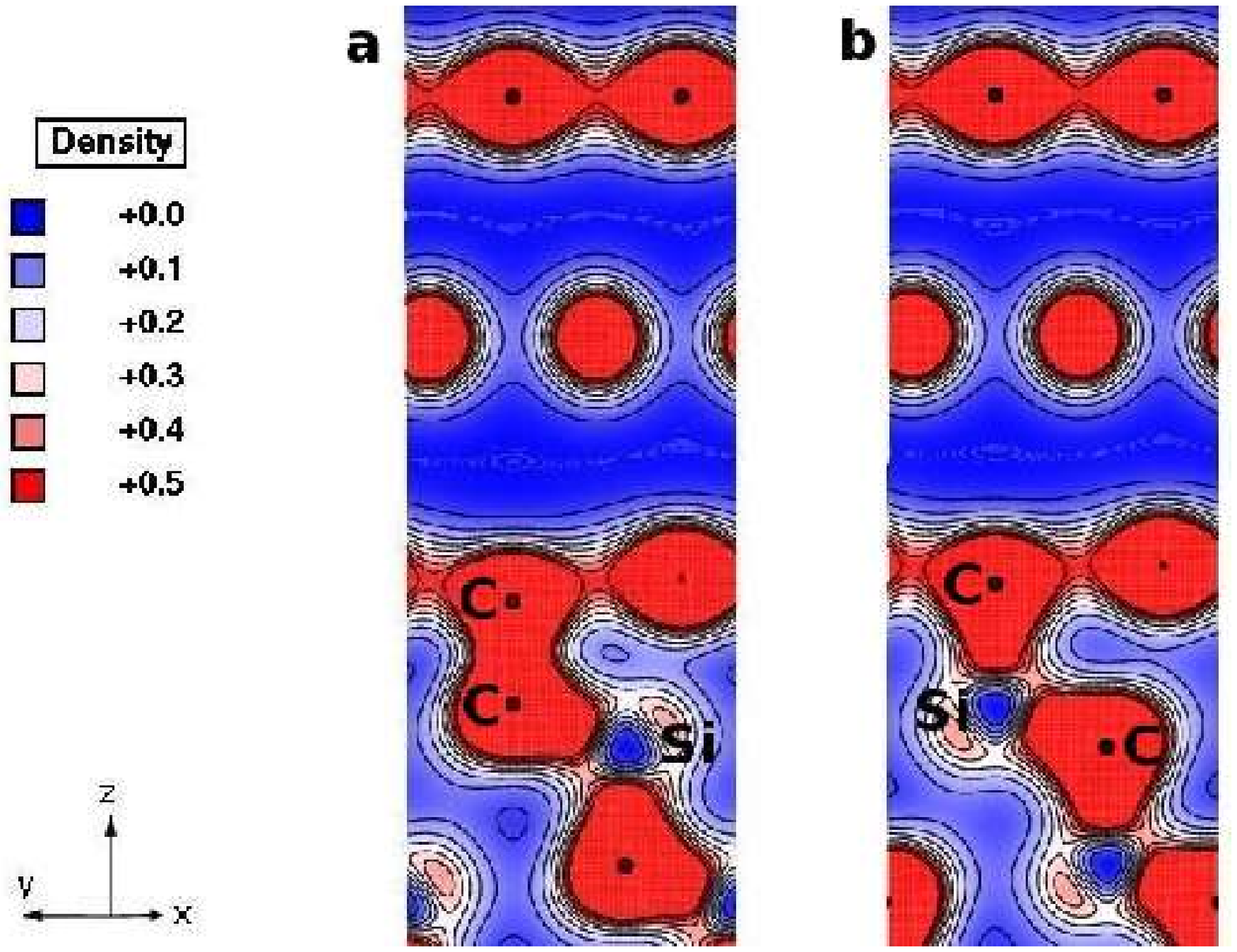}
\caption{(color on line) isocharge density contours along z axis for three C
layers on C- (a) and Si-(b) terminated surface} \label{f.5}
\end{figure}

In Fig.~\ref{f.5}, charge density contours show clear evidence of the existence of 
a covalent bond between the first graphitic C layer (buffer layer) and SiC. Charge density
appears to be more delocalized in the subsequent graphitic planes
as one can expect for graphene like layers.

In conclusion, we have shown that the first C layer on top of a
SiC surface acts as a buffer layer and allows the next graphene
layer to behave electronically like an isolated graphene sheet.
The existence of strong covalent bonds to the first layer is in
agreement with X-ray reflectivity data. The electronic structure
of subsequent graphitic layers depends on the geometry of the
interface and on the number of layers.  Our calculations show clear evidence of a
charge transfer from SiC to the graphene layers that depends on
the interface geometry and results in a doping of these layers. We even show the possible
opening of a gap at the Dirac point in agreement with ARPES
results.
Interface intrinsic defects induce states in the vicinity of the
Fermi level. The interaction of these states with the graphene
derived states depends on interface geometry and may explain the
lower electronic mobility observed on Si-terminated surface.
Because the defect density (i.e. DB states) is even larger for the
actual $6\sqrt{3}\!\times\!6\sqrt{3}R30$ cell, further experiments
and calculations are needed to clarify the role of these states
and their dependence on interface geometry and stacking order in
these systems.

 \subsection*{Acknowledgments}
We thank P.Darancet, D. Mayou, V.Olevano, P.First, W.de Heer for fruitful
discussions. This work is supported by: a computer grant at IDRIS-CNRS and the ACI CIMENT (phynum project); the
National Science Foundation under Grant No. 0404084, by Intel
Research and by the US Department of Energy (DE-FG02-02ER45956).
The Advanced Photon Source is supported by the DOE Office of Basic
Energy Sciences, contract W-31-109-Eng-38.  The $\mu$-CAT beam
line is supported through Ames Lab, operated for the US DOE by
Iowa State University under Contract No.W-7405-Eng-82.  Any
opinions, findings, and conclusions or recommendations expressed
herein are those of the authors and do not necessarily reflect the
views of the research sponsors.

\newpage



\begin{thebibliography}{99}


\bibitem{berger2}
C.Berger et al.,J.Phys. Chem.B {\bf 108}, 19912 (2004)

\bibitem{berger}
C.~Berger et al., Science {\bf 312}, 1191 (2006)

\bibitem{novoselov}
K.~S.~Novoselov et al., Nature {\bf 438}, 197 (2005)

\bibitem{zhang}
Y.~Zhang, Y.-W.~Tan, H.~L.~Stormer,P.~Kim, Nature {\bf 438}, 201 (2005)

\bibitem{edge}
K.~Nakada, M.~Fujita, G.~Dresselhaus, M.~S.~Dresselhaus, Phys. Rev. B {\bf 54}, 17954 (1996)

 \bibitem{edge2}
K.~Wakabayashi, M.~Fujita, H.~Ajiki, M.~Sigrist, Phys. Rev. B {\bf 59}, 8271 (1999)

\bibitem{forbeaux}
I.~Forbeaux, J.~-M.~Themlin, J.~-M.~Debever, Phys. Rev. B {\bf 58}, 16396 (1998)

\bibitem{rollings}
E.Rollings et al., J.Phys.Chem.Sol. {\bf 67}, 2172 (2006)

\bibitem{sadowski}
M.~L.~Sadowski et al., Cond-mat/0605739

\bibitem{ohta}
T.~Ohta, A.~Bostwick, T.~Seyller, K.~Horn, E.~Rotenberg, Science {\bf 313}, 951 (2006)

\bibitem{guinea}
N.M.R.~Peres, F.~Guinea, A.~H.~Castro Neto, Phys. Rev. B {\bf 73}, 125411 (2006)

\bibitem{vasp}
G.~Kresse and J.~Hafner, Phys. Rev. B {\bf 47}, 558 (1993)

\bibitem{pw}
J.~P.~Perdew and Y.~Wang, Phys. Rev. B {\bf33}, 8800 (1986)

\bibitem{uspp}
G.~Kresse and J.~Hafner, J.Phys. Condens. Matter {\bf 6}, 8245 (1994)

\bibitem{incze}
A.~Incze, A.~Pasturel, P.~Peyla, Phys.Rev.B {\bf 66}, 172101 (2002); A.Incze PhD thesis (2002), Grenoble, France

\bibitem{vdW}
The first C layer position is determined by the formation of  strong covalent bonds with the substrate. This is well described by DFT. For subsequent C layers, we changed the interlayer spacing from 2 to 5 \!\AA\! in the system made of 2 C layers on top of a Si terminated SiC surface. The change in total energy is lower than 0.5 $ 10^{-3}$ eV.  The E(d) (d is the C interlayer spacing) curve is similar to the curve calculated for graphite \cite{evc}. 

\bibitem{evc}
N.~Mounet and N.~Marzari, Phys. Rev. B {\bf 71}, 205214 (2005)

\bibitem{x_ray}
J.~Hass, R.~Feng, T.~Li, X.~Li, Z.~Zong, W.A.~de Heer, P.N.~First,
E.H.~Conrad, C.A.~Jeffrey, and C.~Berger, Appl. Phys. Lett. {\bf
89}, 143106 (2006)

\bibitem{X-ray2}
R.~Feng, J.~Hass, J.~Mill\'{a}n, X.~Li, M.~Sprinkle, P.N.~ First
C.~Berger, and E.H.~Conrad (to be published).

\bibitem{Bulk_C}
Y.~Baskin and L.~Mayer, Phys. Rev. {\bf 100} 544, (1955) 

\bibitem{Faults}
R.~E.~Franklin, Acta Cryst. {\bf 4} 253, (1951)

\bibitem{vanBommel75}
A.~J.~Van Bommel, J.~E.~Crombeen, and A.~Van Tooren, Surf. Sci. {\bf 48}, 463 (1975).


\bibitem{mattausch}
This can also be seen in a preliminary calculation by A.~Mattausch and O.~Pankratov published in the Proceedings of ECSCRM 2006.

\bibitem{latil}
S.~Latil, L.~Henrard, Phys. Rev. Lett. {\bf 97}, 036803 (2006)






\end{thebibliography}
\end{document}